\documentclass[12pt]{article}
\usepackage[utf8]{inputenc}
\usepackage[T1]{fontenc}
\usepackage{graphicx}
\usepackage{units}
\usepackage{dsfont}
\usepackage{upgreek}
\usepackage{amsfonts}
\usepackage{amsmath}
\usepackage{authblk}

\begin{document}

\title{Pearcey solitons in curved nonlinear photonic caustic lattices}

\author[]{A. Zannotti}
\author[]{M. Rüschenbaum}
\author[]{C. Denz}
\affil[]{Institute of Applied Physics and Center for Nonlinear Science (CeNoS), University of Münster, 48149 Münster, Germany}
\affil[]{\textit {a.zannotti@uni-muenster.de}}
\date{February 2017}
\maketitle

\begin{abstract}
Controlling artificial Pearcey and swallowtail beams allows realizing caustic lattices in nonlinear photosensitive media at very low light intensities. We examine their functionality as 2D and 3D waveguiding structures, and show the potential of exploiting these lattices as a linear beam splitter, which we name a Pearcey-Y-splitter. For symmetrized Pearcey beams as auto-focusing beams, the formation of solitons in focusing nonlinearity is observed. Our original approach represents the first realization of caustic photonic lattices and can directly be applied in signal processing, microscopy and material lithography.
\end{abstract}


\section{Introduction}

Caustics occur as natural phenomena in light: Light rays refracted and reflected by spherical water droplets form a caustic that mathematically represents a so-called fold catastrophe, and physically appears as a rainbow\cite{Berry1980}. A cusp catastrophe in turn is the origin of light intensity reflection in a cup, and caustic networks form at the bottom of shallow waters.

These processes of stable focusing find their description and hierarchical categorization in the catastrophe theory~\cite{Arnold2003}. Its capability to predict the geometrical structure of the singular mapping of the related ray envelope was extensively studied in the late 70s and 80s~\cite{Berry1976, Berry1979, Kravtsov1983}. It experiences a renaissance nowadays by embedding artificially designed caustics in paraxial beams. In 2007, the most fundamental fold catastrophe was shown to be directly related to the Airy beam~\cite{Siviloglou2007a}, followed by the second-order cusp catastrophe that can be connected to the Pearcey beam in 2012~\cite{Ring2012}. Recently, higher-order catastrophes were demonstrated in swallowtail and butterfly beams owing their names to the corresponding form of caustics~\cite{Zannotti2016}. 

These caustics find importance in their capability to optically represent properties of the dynamics of nonlinear systems in light structures. However, in particular their unique accelerated propagation characteristics make them also highly attractive for many applications. The Airy beam propagates in a transverse invariant way on a parabolic trajectory~\cite{Siviloglou2007a}, the Pearcey beam shows form-invariance and a strong auto-focusing~\cite{Ring2012}, and the swallowtail beam attracts high interest due to its propagation on curved trajectories, whereby different orders of catastrophes are transferred into each other~\cite{Zannotti2017}.

In our contribution, we utilize the trajectories of higher-order caustic beams to optically induce caustic lattices in photosensitive media at low light intensities of a few ten microwatts laser power. Even at these low intensities, the trajectories of higher orders form high-intensity structures that are capable to realize defined refractive index lattices with outstanding guiding features. Thus, these photonic structures are especially suited to act as waveguides with a rich diversity of light guiding paths. Here we demonstrate guiding along transverse, two-dimensional (2D) quasi cubic and three-dimensional (3D) disposed curved waveguides, as well as their functionality as optical splitters. Finally, taking advantage of the strong auto-focusing of Pearcey beams, we demonstrate the formation of a novel and up to now unobserved solitary wave in a nonlinear photorefractive crystal which we name Pearcey soliton.

\section{Caustic light and experimental setup}

The fundamental light structures that are used in the following to realize photonic lattices are a Pearcey beam
\begin{equation}
\text{Pe}\left(x,y\right) = \int_\mathds{R}{\text{exp}\left[\text{i}\left(s^4 + \frac{y}{y_0}s^2 + \frac{x}{x_0}s\right)\right]\text{d}s},
\label{eq:PearceyBeam}
\end{equation}
and a swallowtail beam
\begin{equation}
\text{Sw}\left(x,y\right) = \int_\mathds{R}{\text{exp}\left[\text{i}\left(s^5 + \frac{y}{y_0}s^2 + \frac{x}{x_0}s\right)\right]\text{d}s},
\label{eq:SwallowtailBeam}
\end{equation}
where, in terms of catastrophe theory, $s$ is the so-called state parameter of the nonlinear system, and $x, y$ are its control parameters. $x_0, y_0$ represent the beams' structure sizes, and are connected in general, as for the case of the Gaussian beam, with the characteristic Rayleigh length $z_e = 2ky_0^2$. In case of the auto-focusing Pearcey beam, the focal point is located at $z = z_e$, depending on the wave number $k = 2\pi/\lambda$, and the wavelength $\lambda$~\cite{Ring2012, Zannotti2017}.

\begin{figure}
\centering
\includegraphics[width=.75\columnwidth]{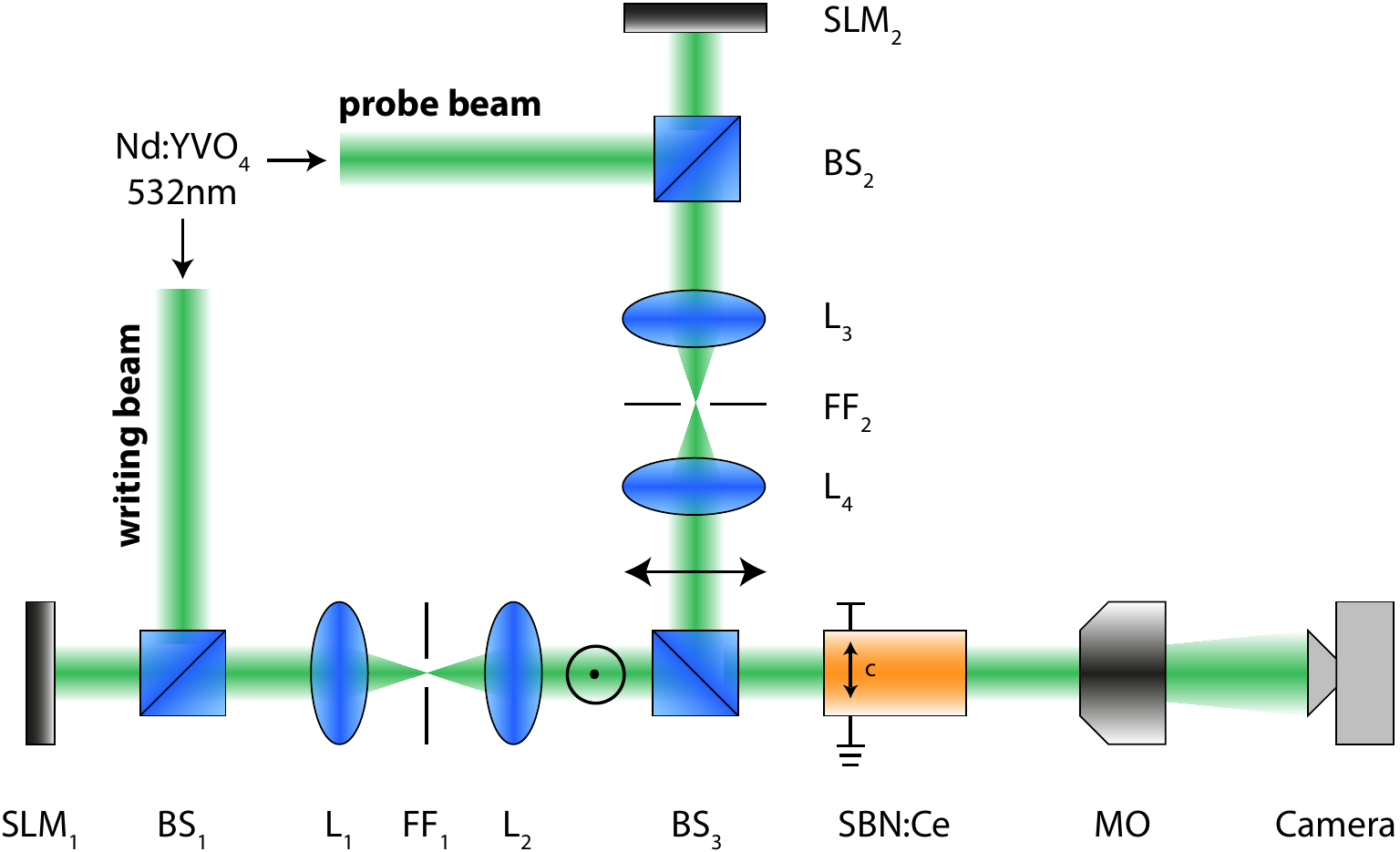}
\caption{Experimental setup. BS: beam splitter, FF: Fourier filter, L: lens, MO: microscope objective, SLM: spatial light modulator.}
\label{fig:ExperimentalSetup}
\end{figure}

In order to optically induce curved refractive index channels that can be exploited as innovative waveguides in photorefractive media, we use a setup as sketched in Fig.~\ref{fig:ExperimentalSetup}, and along the description in~\cite{Diebel2014, Diebel2015}. The light field of a frequency-doubled Nd:YVO$_4$ laser beam with $\lambda = \unit[532]{nm}$ is split in two beams: a writing and a probe beam. Both are modulated in amplitude and phase with phase-only spatial light modulators (SLM$_{1,2}$). Therefore, appropriate Fourier filters (FF) are necessary~\cite{Davis1999}. An ordinary polarized writing beam with typical powers in the order of a few $\unit[10]{\upmu W}$ illuminates a biased photorefractive SBN:Ce crystal with bulk refractive indices of $n_\text{o} = 2.358$ and $n_\text{e} = 2.325$, and electro-optic coefficients of $r_{13} = \unit[47]{V/pm}$ and $r_{33} = \unit[237]{V/pm}$, respectively. Its geometrical dimensions are $\unit[5 \times 5 \times 20]{mm^3}$. The direction of propagation is along the $z$-axis parallel to the long axis of the crystal, and its optical $c = x$-axis is perpendicular to this. The writing light forms accordingly a refractive index structure that represents the artificial caustic lattices. Subsequently, extraordinary polarized Gaussian probe beams are guided in the photonic structures. A camera either images the transverse intensity distribution at the front or back face of the crystal.

\section{Waveguiding in photonic caustic lattices}

\begin{figure}
\centering
\includegraphics[width=.75\columnwidth]{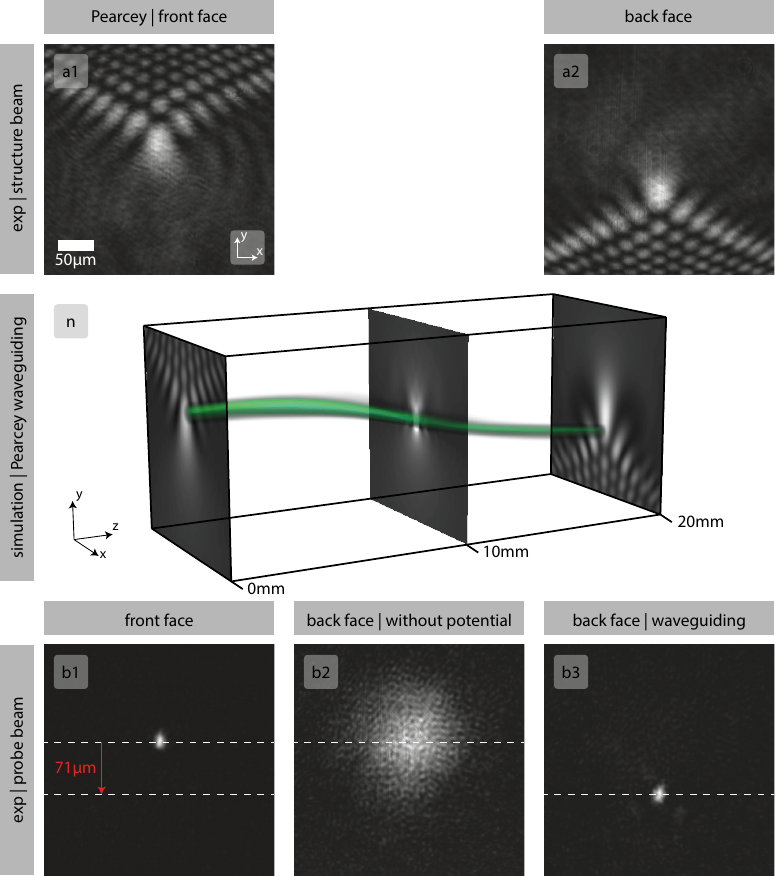}
\caption{2D waveguiding in a Pearcey lattice. (a) Characterization of the structure beam. (b) A Gaussian probe beam is guided along the quasi-cubic waveguide. (n) Numerical simulation of a linearly propagating Gaussian (green) in the Pearcey lattice (gray).}
\label{fig:PearceyWaveguide}
\end{figure}

Fig.~\ref{fig:PearceyWaveguide} shows the creation of a Pearcey photonic lattice and linear waveguiding along a quasi cubic path. Here, we utilize a standard Pearcey beam as stated in Eq.~\eqref{eq:PearceyBeam} with a structure size of $x_0 = y_0 = \unit[13.4]{\upmu m}$, so that the focus is located at half of the length of the crystal at $z = z_e = \unit[10]{mm}$. The Pearcey beam at the entrance and output face of the crystal is shown in Fig.~\ref{fig:PearceyWaveguide}(a) when linearly propagating in the medium.

Applying a high voltage of $E_\text{ext} = \unit[2000]{V/cm}$ and a beam power of $P = \unit[50]{\upmu W}$ for $t = \unit[40]{s}$, we nonlinearly induce a refractive index lattice according to this writing beam that represents a waveguide due to nonlinear refractive index enhancement by the Pearcey structure. Figs.~\ref{fig:PearceyWaveguide}(b) show an extraordinarily polarized Gaussian beam with an initial beam waist of $w_0 = \unit[10]{\upmu m}$ (FWHM) at the front~(b1) and back face of the crystal for its linear propagation through a homogeneous crystal~(b2) and the Pearcey lattice~(b3). We clearly identify a transverse shift of the Gaussian spot of $\approx\unit[71]{\upmu m}$ guiding along the lattice. Moderate losses are apparent and can be estimated by the ratio of the output ($P_2$) and input ($P_1$) powers $Q = 10\lg \frac{P_2}{P_1} \approx \unit[-4.0]{dB}$. According to Eq.~\eqref{eq:PearceyBeam} and the analytical description of the Pearcey beam propagation~\cite{Ring2012}, the trajectory of the 2D waveguide establishes only in the $y$-$z$-plane. Shown in Fig.~\ref{fig:PearceyWaveguide}(n), we realized numerical simulations on the basis of~\cite{Diebel2015} that give detailed insight in the propagation path inside the crystal. The Pearcey lattice is shown by three slices at characteristic positions at the front face ($z = \unit[0]{mm}$), focal plane ($z = \unit[10]{mm}$), and back face ($z = \unit[20]{mm}$), while the trajectory of the guided Gaussian beam is indicated in green. 

\begin{figure}
\centering
\includegraphics[width=.75\columnwidth]{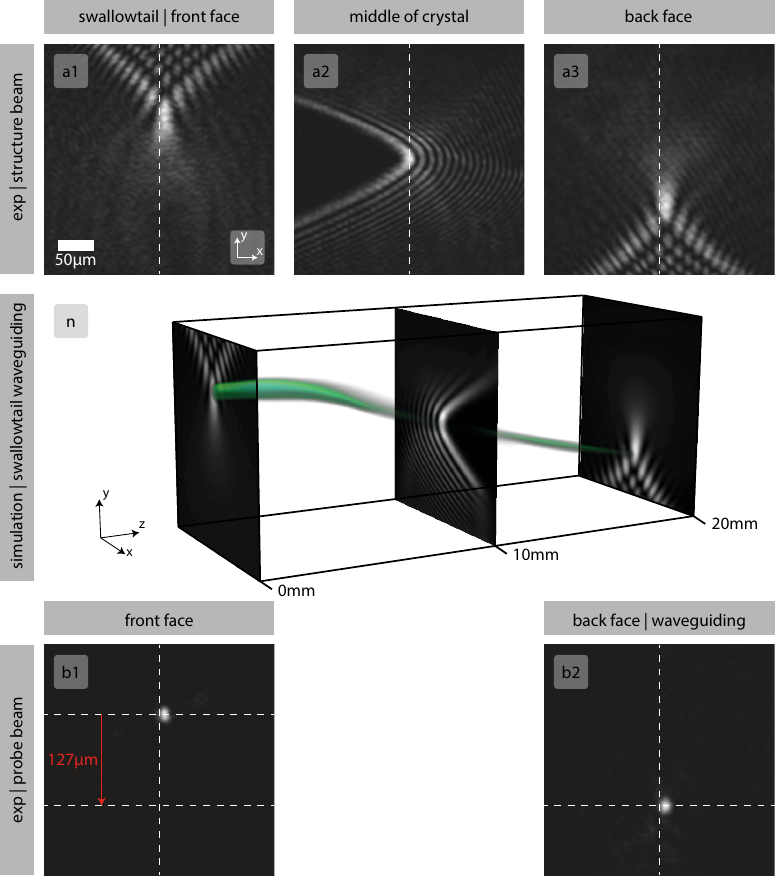}
\caption{Waveguiding on a 3D path in a swallowtail lattice. (a) The numerically pre-calculated swallowtail beam serves as structure beam. (b) A Gaussian probe beam propagates on a 3D curved path. (n) Numerical simulation of a linearly propagating Gaussian (green) in the swallowtail lattice (gray).}
\label{fig:SwallowtailWaveguide}
\end{figure}

The potential of caustic beams for refractive index modifications lies in their manifold of individual high-intensity trajectories. We demonstrate exemplarily refractive index formation by a standard swallowtail beam according to Eq.~\eqref{eq:SwallowtailBeam}, whose main intensity propagates on a 3D curved path. It was shown~\cite{Zannotti2017} that the swallowtail beam exhibits point symmetry with respect to the $z = 0$ transverse plane. Consequently, we implement a swallowtail beam with $x_0 = y_0 = \unit[4.2]{\upmu m}$, whose symmetry plane is positioned in the middle of the crystal at $z = \unit[10]{mm}$. In Fig.~\ref{fig:SwallowtailWaveguide}(a), the beam is shown at the front (a1) and the back face (a3) as well as at the half-length (a2) of the crystal. We inscribe a swallowtail lattice in the nonlinear medium according to the procedure described above. The parameters were the same for the formation of the refractive index structure. Figs.~\ref{fig:SwallowtailWaveguide}(a) clearly show a small shift in $x$-direction when following the path of the main intensity. The guiding in $y$-direction is obvious when comparing input and output of a Gaussian beam. It is imaged in \ref{fig:SwallowtailWaveguide}(b) with an initial beam waist of $w_0 = \unit[10]{\upmu m}$ and propagates linearly following the induced curved waveguide. During propagation, the beam size stays almost constant. Our numerical simulation of wave guiding in swallowtail lattices in Fig.~\ref{fig:SwallowtailWaveguide}(n) is in very good agreement with the experiment. Three slices indicate the formation of the lattice according to the swallowtail light structure, and the path of the Gaussian probe beam is highlighted in green. The experimentally measured leakage of energy is moderate, apparent at the ratio of output and input power of $Q \approx \unit[-2.1]{dB}$. 

The two examples selected for the formation of curved caustic lattices clearly demonstrate the extensive potential of all caustic beams to realize innovative wave guiding. Moreover, combining different orders results in multi-focal beams that paves the way to realize a number of diverse refractive index structures and thus waveguides in a single formation process in parallel. 

\begin{figure}
\centering
\includegraphics[width=.75\columnwidth]{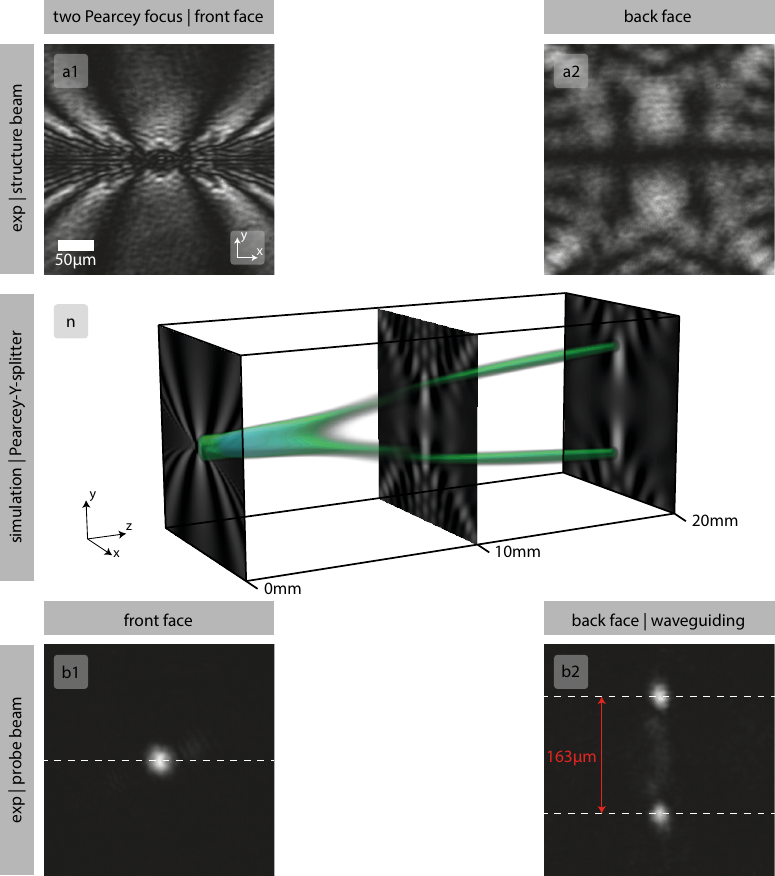}
\caption{A Pearcey-Y-splitter. (a) Two mirrored out-of-phase Pearcey beams form the structure beam. (b) A Gaussian probe beam is split and guided in two separated channels. (n) Numerical simulation of a linearly propagating Gaussian (green) in the Pearcey-Y-splitter (gray).}
\label{fig:PearceySplitter}
\end{figure}

Out of the many attractive examples for these wave guiding applications that can be implemented with our approach, we have selected a Y-splitter capable of splitting and merging optical signals which we realize exploiting the Pearcey beam structure. It is shown in Fig.~\ref{fig:PearceySplitter}. Therefore, we create an initial field distribution $E\left(x,y,0\right)$ synthesized by two out-of-phase Pearcey beams and shifted their focal positions to the entrance face of the crystal:
\begin{equation}
E\left(x,y,0\right) = \text{Pe}\left(x,y,z_e\right) + e^{\text{i}\pi} \cdot \text{Pe}\left(x,-y,z_e\right).
\label{eq:PearceySplitter}
\end{equation}
Here, $x_0 = y_0 = \unit[18.9]{\upmu m}$. The composed structure is imaged at the entrance and the output crystal face in Fig.~\ref{fig:PearceySplitter}(a). We benefit from the auto-focusing feature of this beam and its simultaneous form-invariance.  The resulting composed structure is characterized by two high-intensity maxima in the initial plane that drift away from each other on quasi cubic paths during propagation in $z$-direction. Due to the cuspoid form of each of the two beams that point in opposite directions, the advantage is that the focal spots are not interfering and can independently induce curved waveguides. The induction time was $t = \unit[21]{s}$. Figs.~\ref{fig:PearceySplitter}(b) show the Gaussian probe beam with a beam waist of $w_0 = \unit[20]{\upmu m}$, and its output transverse intensity distribution. For this linear propagation of the probe beam, we demonstrate the functionality of the Y-splitter by showing two symmetric and distinctive output signals. Our numerical simulation of the probe beam propagation in the Pearcey-Y-splitter structure is shown in Fig.~\ref{fig:PearceySplitter}(n) supporting clearly this functionality. 

\section{Formation of a Pearcey soliton}
\begin{figure}
\centering
\includegraphics[width=.75\columnwidth]{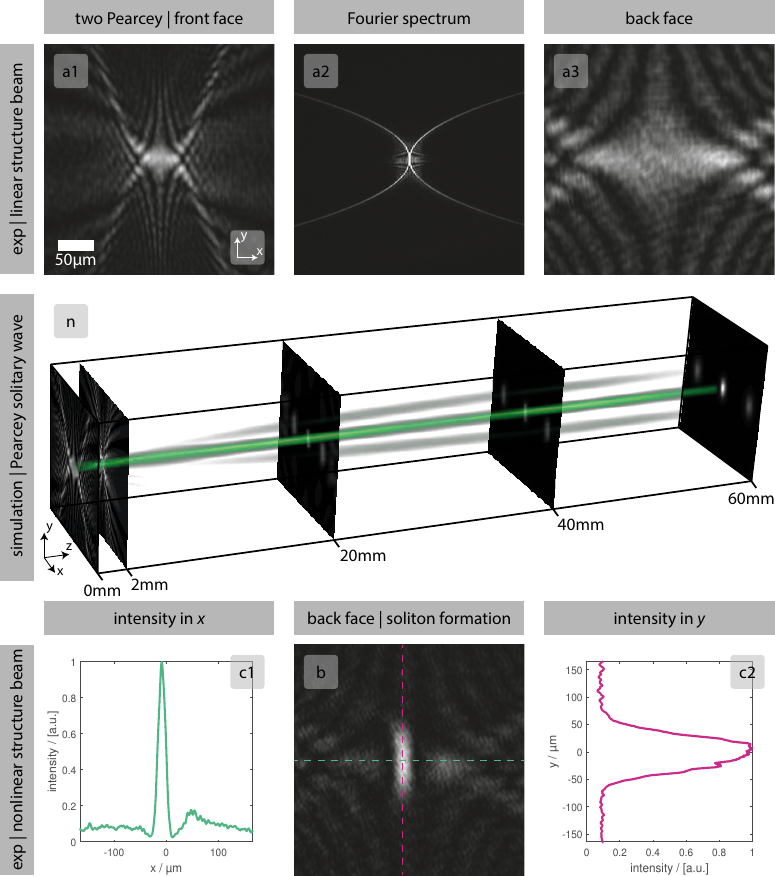}
\caption{Formation of a Pearcey soliton. (a) Transverse intensity distribution of the initial field at the front face in real (a1) and Fourier space (a2) as well as linearly propagated at the back face of the crystal (a3). (b) Transverse intensity pattern of the realized Pearcey soliton. (c1), (c2) Quantification of the transverse intensity distribution along the $x$- and $y$-direction, respectively.}
\label{fig:PearceySoliton}
\end{figure}

There are several properties the Pearcey beam exhibits that make this caustic beam highly suited for nonlinear wave propagation: Pearcey beams perform a strong auto-focusing, and their maximum intensity can easily be tailored to meet the threshold of nonlinear self-focusing where the nonlinearity compensates diffraction which is the mandatory requirement for nonlinear spatial soliton formation. However, a single Pearcey beam induced as refractive index lattice is not able to create spatial solitons, due to the asymmetry it exhibits along the $y$ direction, which prevents balance of diffraction and nonlinear self-focusing due to a stronger diffraction contribution. But exploiting the form-invariance and flipping of one of the Pearcey beams by changing  $y$ to $-y$ at its focal position $z = z_e$ leads to a field distribution that features the formation of a spatial soliton when superimposing symmetrically two regarding $y$ mirrored Pearcey beams, as shown in Fig.~\ref{fig:PearceySoliton}. Additionally, this light field does not posses interfering side lopes during propagation, nor before neither behind the focal position $z_e$. Especially, this yields a more homogeneous transverse light distribution without disturbances or excitation of oscillations as it has been identified to be prerequisites for stable soliton formation in e.~g.~Airy lattices~\cite{Diebel2015}.

Thus, we decided to realize a soliton-supporting refractive index structure by superimposing two in-phase Pearcey beams creating constructive high-intensity interference at $(x,y,z) = (0,0,z_e)$. Since both Pearcey beams must exactly overlap at this focal point, but shift in $y$ direction during propagation~\cite{Ring2012}, we need to precalculate an adequate initial field distribution $E\left(x,y,0\right)$ with transversely displaced Pearcey beams. This can be realized by
\begin{equation}
E\left(x,y,0\right) = \text{Pe}\left(x,y-\frac{y_0^3}{x_0^2},0\right) + \text{Pe}\left(x,-y-\frac{y_0^3}{x_0^2},0\right).
\label{eq:PearceySoliton}
\end{equation}
Note the dependence of the initial displacement $y_0^3/x_0^2$ on the focal position $z = z_e = 2ky_0^2$.

Fig.~\ref{fig:PearceySoliton}(a1) shows the experimentally obtained initial intensity distribution of the ordinarily polarized writing beam imaged at the front face of the homogeneous crystal, where $z_e = \unit[2]{mm}$, thus $x_0 = y_0 = \unit[6]{\upmu m}$. By temporarily placing an additional lens behind the crystal, we experimentally accessed the Fourier spectrum of the writing beam in Fig.~\ref{fig:PearceySoliton}(a2), that consists of two parabola~\cite{Ring2012}. Fig.~\ref{fig:PearceySoliton}(a3) shows the linearly propagated writing beam at the back face of the crystal. Subsequently, we studied the conditions for soliton formation. The essential properties for soliton formation are an electric bias field of $E_\text{ext} = \unit[2000]{V/cm}$ and a beam power of $P \approx \unit[30]{\upmu W}$ to induce nonlinear self-focusing. The soliton starts to form after the focal position at $z = \unit[2]{mm}$, and propagates with a spatially invariant transverse beam profile from a position of $z \approx \unit[3]{mm}$. The experimentally obtained output intensity distribution clearly demonstrates this novel soliton which we name 'Pearcey soliton'. It is shown in Fig.~\ref{fig:PearceySoliton}(b). The well-known effect of an anisotropic transverse shift of a soliton in a photorefractive medium~\cite{Diebel2015} is also observed in Fig.~\ref{fig:PearceySoliton}(b). It can be attributed to the additional diffusion-based charge transport component that acts additionally to the external field-based component in SBN:Ce 60~\cite{Petter1999}. In order to proof soliton formation, we quantify the transverse intensity distribution along the x- and y-direction in Figs.~\ref{fig:PearceySoliton}(c1) and (c2), respectively. Our corresponding numerical simulation in Fig.~\ref{fig:PearceySoliton}(n) provides insight into the dynamic soliton formation in the nonlinear medium. In order to numerically verify and clearly support the soliton-forming capability of the Pearcey writing beam, we extend the propagation distance to $\unit[60]{mm}$, which is not only three times the crystal length but $\approx 29$ times the Rayleigh length $z_e$ of the Pearcey writing beam. In Fig.~\ref{fig:PearceySoliton}(n), the initial beam shape ($z = \unit[0]{mm}$), the focal plane ($z = \unit[2]{mm}$), the output plane of our crystal ($z = \unit[20]{mm}$), and the longest investigated distance ($z = \unit[60]{mm}$) are highlighted as transverse slices. In green, we indicate the beam intensity distribution during propagation.

\section{Conclusion}
To summarize, we demonstrated a variety of caustic lattices created by low light intensities in photosensitive photorefractive materials with particular emphasis on their potential to act as curved waveguides with a rich manifold of diverse trajectories in 2D and 3D. Exemplarily, we showed caustic Pearcey and swallowtail beams to artificially create photonic lattices. Subsequently, we probed them with Gaussian beams in the linear regime. Caustic beams are especially suited to realize lattices for waveguiding due to their auto-focusing and curved paths of main intensities during propagation. Additionally, by creating a Pearcey-Y-splitter and prove its functionality, we pave the way to future information processing applications. We want to emphasize the huge potential that especially combined caustic beam structures exhibit in their focus regions for material processing or optical micro manipulations at low intensities. 

Exploiting the regime where nonlinear self-focusing and diffraction of Pearcey beams balance, we were able to realize for the first time soliton formation in tailored Pearcey lattices. For this purpose, we created composite Pearcey-based, caustic light fields consisting of two Pearcey beams propagating towards each other, and thus tailored the focus structure. We were able to demonstrate a novel class of solitons which we name Pearcey solitons that was observed to be stable for experimental distances of $\unit[20]{mm}$, backed up by numerical simulations. 

With our work, we pioneered photonic caustic lattices and pave the way to apply this approach to a manifold of applications ranging from linear waveguiding to nonlinear soliton formation. Since this beam class provides diverse light structures that each excels by unique propagation properties which furthermore can be combined to more complex light fields, we are convinced to have presented an innovative toolbox with far-reaching applications in material processing, micro-manipulation, signal processing, and microscopy.

\section{Acknowledgement}
We acknowledge helpful discussion with Falko Diebel. Part of this work was supported especially by his advices with respect to the optical experiments.

\bibliographystyle{unsrt}

\end{document}